\documentclass[11pt, letter]{article}
\usepackage{jheppub}

\def\be{\begin{equation}}
\def\ee{\end{equation}}
\newcommand{\kah}{K\"ahler}
\renewcommand{\to}{\longrightarrow}

\usepackage[dvipsnames]{xcolor}

\title{Quadratic curvature corrections to stringy effective actions and the absence of de Sitter vacua}

\author[1,2]{Francesc Cunillera,}
\author[3]{William~T.~Emond,}
\author[4]{Antoine Leh\'{e}bel}
\author[1,2]{and~Antonio Padilla}

\affiliation[1]{School of Physics and Astronomy, University of Nottingham, University Park, Nottingham NG7 2RD, United Kingdom}
\affiliation[2]{Nottingham Centre of Gravity, University of Nottingham, Nottingham NG7 2RD, UK}
\affiliation[3]{CEICO, Institute of Physics of the Czech Academy of Sciences, Na Slovance 2, 182 21 Praha 8, Czech Republic}
\affiliation[4]{Centro de Astrof\'{i}sica e Gravita\c{c}\~{a}o - CENTRA, Departamento de F\'{i}sica, Instituto Superior T\'{e}cnico - IST, Universidade de Lisboa - UL, Av. Rovisco Pais 1, 1049-001 Lisboa, Portugal}

\emailAdd{william.emond@fzu.cz}
\emailAdd{francesc.cunillera@nottingham.ac.uk}
\emailAdd{antoine.lehebel@tecnico.ulisboa.pt}
\emailAdd{antonio.padilla@nottingham.ac.uk}
\abstract{We investigate the combined effect of fluxes and higher-order curvature corrections, in the form of the Gauss-Bonnet term, on the existence of de Sitter vacua in a heterotic string inspired framework, compactified on spheres and tori. We first gain some intuition on the effects of these corrections by studying a perturbative expansion in the small Gauss-Bonnet coupling. Then, for choices of potential closer to the string theory predictions, we show that the inclusion of quadratic curvature corrections actually reduces the parametric likelihood of  de Sitter solutions.
} 

\begin{document}
\maketitle

\section{Introduction}

A slew of observations, from supernova \cite{SN1,SN2} to the cosmic microwave background \cite{CMB},  point to a standard model of cosmology in which the universe  was dominated by radiation at early times, then by matter, before entering the current phase of dark energy domination.  The microscopic origin of dark energy has not been established  although on cosmological scales we know that it behaves like a fluid whose equation of state is very close to $\omega_\text{DE} = -1$ \cite{derev}.  This leads to a near constant energy density, consistent with a cosmological constant \cite{dS} or a slowly rolling quintessence field \cite{quin1a,quin1b,quin2}.

 The existence of a (meta)stable de Sitter vacuum is one of the most important open questions in string theory. Tree-level classical supergravities yield a flat potential for a number of moduli fields. Stabilising these fields generically requires the introduction of perturbative and non-perturbative corrections to the tree-level action. These corrections are fundamental to supergravity model building and all known de Sitter constructions make use of them, such as in  KKLT \cite{Kachru:2003aw} and LVS models \cite{hep-th/0502058}. Obtaining a systematic understanding of the hierarchy of these corrections is a highly non-trivial task. A reason behind the difficulty of the task at hand is that supergravity inherits two perturbative parameters from string theory: {the Regge slope} $\alpha'$ and the string coupling $g_\text{s}$. In writing down quantum corrections to the tree-level action, we will inevitably encounter terms of order $\mathcal{O}(\alpha'^p g_\text{s}^q)$, for some $p,~q>0$. As a result, it can sometimes become unclear which terms enter at which order in perturbation theory. For example, comparing terms $\mathcal{O}(\alpha'^3 g_\text{s}^2)$ and $\mathcal{O}(\alpha'^2 g_\text{s}^3)$ can be hard. A very important part of ongoing supergravity research is to systematically analyse these corrections at all orders and ensure that \kah moduli stabilisation is not spoiled in the bulk of moduli space (see for example \cite{Cicoli:2007xp, Cicoli:2008va, Kachru:2019dvo, AbdusSalam:2020ywo, Cicoli:2021rub}). Furthermore, since we lack a complete picture of non-perturbative supergravity, it is complicated to define frameworks where computations can be carried out under complete control. Defining {\it complete control} is itself a hotly debated topic in string phenomenology. The difficulty of finding de Sitter vacua could make the idea of a dynamical model of dark energy in the bulk of moduli space more appealing. Such models could also offer a solution to the cosmological coincidence problem \cite{coin}.  However, it should be noted that engineering a viable quintessence model building has many difficulties \cite{Garg1, ivonne, no_runaway, Hebecker, us}, and it is likely that quintessence could be even harder to obtain than a true de Sitter vacuum \cite{no_runaway, us}.

The lack of a completely satisfactory answer to the challenges to de Sitter model building within supergravity presented above led to the creation of the so-called {\it Swampland programme} (see \cite{Palti:2019pca} for an in-depth review). Its core goal is to differentiate between low energy effective field theories that have a consistent completion within a prospective quantum gravity UV theory and those that do not. In doing so, the Swampland programme has formed a {\it web of conjectures} that attempt to delineate the space between the consistent and inconsistent theories. In \cite{Vafa}, the idea that not every low-energy effective field theory can descend from a consistent string theory compactification was put forward. Critically, the arguments of \cite{Vafa} are only strictly valid on the boundary of moduli space. Another conjecture of particular cosmological interest is the so-called de Sitter conjecture \cite{Obied, Garg:2018reu} (see also \cite{Danielsson}), which claims that (meta)stable de Sitter vacua are incompatible with string theory and further tries to constrain the type of potentials that can be derived from string theory. If one takes the de Sitter conjecture to be true, our late time Universe would have to be described by a dynamical model of dark energy at the boundary of moduli space. In \cite{Garg1, ivonne, no_runaway}, it was further shown that no slow roll regime can appear in the boundary of moduli space, implying that a quintessence regime is not viable either. This suggests a possible disconnect between the phenomenological conjectures of the Swampland programme and observational cosmology. 

Many of the arguments we have just described rely on compactifications of string theory on Calabi-Yau manifolds, originally chosen to preserve some supersymmetry in the external space. Of course, to recover de Sitter space, supersymmetry must be broken completely\footnote{We note that supersymmetry may be realised non-linearly on de Sitter space whenever the symmetry is broken spontaneously, as in \cite{Dudas:2015eha, Bergshoeff:2015tra, Bandos:2015xnf, Nagy:2019ywi, Bansal:2020krz}}, so it is worth exploring string compactifications on manifolds that are not Calabi-Yau, at least in the rare cases where one can solve the equations and retain  some calculational control. To this end, in \cite{Montero}, the authors  took a generic class of string inspired models and studied compactifications on a $p$-sphere with internal $p$-form fluxes. They identified a parametric bound (henceforth known as the Montero-van Riet-Venken (MvRV) bound), that needed to hold for de Sitter solutions to exist. The MvRV bound corresponds to a remarkably simple inequality between the slope of the potential and the derivative of a $p$-form gauge coupling. A number of stringy models were studied where it could be explicitly shown that the inequality is not satisfied, ruling out de Sitter solutions in the most realistic set-ups. 

In this paper, we expand on the work of \cite{Montero} by introducing higher order curvature corrections in the form of a Gauss-Bonnet contribution to the action, as well as generalising the set-up to contain internal or external fluxes with spherical or toroidal internal geometries. The goal is to establish whether the curvature corrections  strengthen or weaken the MvRV bound of \cite{Montero} for existence of de Sitter vacua. We will consider actions of the following form:
\begin{equation}
	S=\displaystyle \int{\mathrm{d}^Dx\sqrt{-g}\left[\dfrac{R}{2\kappa_D^2}-\dfrac12 (\partial\varphi)^2-V(\varphi)-{g(\varphi)}\left|H_k\right|^2+f(\varphi)\mathcal{G}\right]},
	\label{eq:action}
\end{equation}
and take our ansatz manifold to be the warped product $\mathcal{M}=\Sigma_n\rtimes_\rho\Sigma_p$ with line element
\begin{equation}
\label{eq:metansatz}
	\mathrm{d}s^2=g_{AB}\mathrm{d}x^A\mathrm{d}x^B=e^{2\alpha \rho}\tilde{g}_{\mu\nu}\mathrm{d}\tilde{x}^\mu\mathrm{d}\tilde{x}^\nu+e^{2\beta \rho}\hat{g}_{ab}\mathrm{d}\hat{x}^a\mathrm{d}\hat{x}^b,
\end{equation}
where $\alpha$ and $\beta$ are constants, capital Latin indices refer to the $D$-dimensional spacetime, Greek indices refer to the external spacetime of dimension $n$ with metric $\tilde{g}_{\mu\nu}$, lower case Latin indices refer to the internal manifold of dimension $p$ with metric $\hat{g}_{ab}$, and $\rho$ is a scalar that varies over the external spacetime only. $H_k$ is a top-form that is either purely external ($k=n=D-p$) or purely internal ($k=p$) since there is no preference for one or the other at the level of the effective field theory.  We will consider $\Sigma_n$ to be either de Sitter $dS_n$ or Minkowski $\mathbb{R}^{1,n-1}$, while $\Sigma_p$ will be either  a sphere $S^p$ or a torus $T^p$.

We initially adopt a perturbative approach, working to leading order in the Gauss-Bonnet coupling, $f(\varphi)$.  In this limit, we find  that the MvRV bound obtained in \cite{Montero}  can be marginally violated and de Sitter solutions can still exist.  This is true as long as a second parametric bound holds on the gradient of $f(\varphi)$. Unfortunately, this second bound does {\bf not} hold for choices of $f(\varphi)$ and  $g(\varphi)$ best motivated from string theory. To reinforce this latter point, we extend our analysis beyond perturbation theory, this time specialising to the string motivated choices for $f(\varphi)$ and $g(\varphi)$.  Consistent with our perturbative analysis, we find that the MvRV bound is still a necessary condition for the existence of de Sitter solutions. However, it is no longer sufficient, suggesting that for string motivated potentials, the higher curvature corrections make it parametrically harder, not easier to find de Sitter vacua. 

The structure of the paper is as follows. In section \ref{sec:intro_het}, we review the results of \cite{Montero} and motivate the introduction of the Gauss-Bonnet term as a higher order curvature correction within the framework of heterotic strings. In section \ref{sec:genpot}, we gain some insight into the effects of the Gauss-Bonnet correction by performing a small coupling expansion. Then, in section \ref{sec:specifpot}, we focus on specific string-inspired potentials and discuss in detail what conditions are necessary and sufficient for the existence of solutions. We conclude with a discussion in section \ref{sec:discussion}.


\section{Review of flux compactifications of the heterotic string action \& stability conditions}
\label{sec:intro_het}

In \cite{Montero}, the authors considered the existence and stability of $dS_{n}\times S^p$ flux compactifications. The action of \cite{Montero} can be cast as \eqref{eq:action} with $f(\varphi)=0$ and with
$k=p$, so that $H_k$ is a top-form living in the internal space, providing the flux that aims to stabilise the potential for the scalar $\varphi$, and prevent the runaway. In this particular case, the coefficients $\alpha$ and $\beta$ which parametrise the compactification are chosen to be
\be \label{albe}
\alpha^2=\frac{p}{2(n-2)(n+p-2)}\ , \qquad \beta=-\frac{n-2}{p}\alpha\ ,
\ee
so that the $n$ dimensional effective action obtained after dimensional reduction can be expressed in Einstein frame, with a canonical kinetic term for the breathing mode. The effective potential for $\varphi$ and $\rho$ in the external spacetime is found to be\footnote{Up to an overall internal volume factor which can be reabsorbed in a Weyl redefinition of the four-dimensional metric.}
\be
V_\text{eff}=-\dfrac{p(p-1)}{2\kappa_D^2} e^{2(\alpha-\beta)\rho}+ {Q^2}e^{2(n-1)\alpha \rho}g(\varphi)+V(\varphi)e^{2\alpha\rho}\ .
\ee
As a result, for constant $\varphi$ and $\rho$, the metric field equations in the external $n$ dimensional space take the form $\tilde{G}_{\mu\nu}=-V_\text{eff}\,\tilde{g}_{\mu\nu}$; $Q$ is the magnetic charge associated with the internal top-form, fixed by the equations of motion to be
\be
Q^2=-e^{-2\alpha(n-1)\rho}\frac{V'}{g'}\ ,
\ee
where the primes denote derivative with respect to $\varphi$. Demanding that the potential is positive at its minimum leads to the  MvRV constraint
\be
(p-1)\left|\frac{V'}{V}\right|\leq\left|\frac{g'}{g}\right|\qquad \text{or~equivalently} \qquad  (p-1)\left|\frac{V}{g}\right|\geq e^{-2\alpha(D-p-1)\rho}\frac{1}{Q^2}. \label{eq:constraint}
\ee
To check the stability of such solutions, one must specify the form of the scalar potential $V$ and the coupling between the top-form and the potential $g$, while also stabilising the scalar. In \cite{Montero}, the authors consider a number of string-motivated scenarios with the key takeaway being the instability of the de Sitter minimum. For illustrative purposes, we present below the case where $V(\varphi)=V_0 e^{-\gamma \varphi}$ and $g(\varphi)=e^{\delta \varphi}$, for $\gamma, \delta>0$.  The Hessian of the potential along the $\varphi,\rho$ directions is given by 
\be
\mathcal{H} = {\partial^2 V_{\text{eff}}\over \partial \phi^i \partial \phi^j} \ , \qquad \phi^i=\{\varphi,\rho\}\ .
\ee
and its corresponding determinant by
\be
\det(\mathcal{H})=-\frac{(n-2)[\delta+\gamma(n-1)][\gamma(1-p)+\delta]\delta e^{2\delta \varphi + 2(n-1)\rho}}{4\gamma p}Q^4.
\ee
Demanding that \eqref{eq:constraint} holds at the minimum ---~or in other words $(p-1)\gamma \leq \delta$~--- fixes the determinant of the Hessian to be negative whenever $n>2$. This means all the corresponding solutions are  unstable. The only loophole to the above discussion is the case where the number of external dimensions $n=2$. In this case, a specific analysis is required
and one may obtain a meta-stable de Sitter  solution while satisfying \eqref{eq:constraint}. 

As mentioned in the introduction, here we expand upon the above obstruction by considering the effects of a Gauss-Bonnet term in the action \eqref{eq:action}. Let us now review the origin of this term. Our starting point  is the effective action presented in  \cite{Campbell}
\be
S=\int \text{d}^D x \sqrt{\left|g\right|}\left[{R\over 2\kappa_D^2}-\frac{1}{2}(\partial\varphi)^2-V(\varphi)-g(\varphi)\left|H_3\right|^2+f(\varphi)\mathcal{G}\right]\ , \label{eq:EA_camp}
\ee
with $H_3=\text{d}B_2+\omega_\text{L}-\omega_\text{Y}$ where $B_2$ is the Kalb-Ramond field, $\omega_\text{L}$ is the Lorentz spin connection, $\omega_\text{Y}$ is the Yang-Mills gauge connection, and with
\be
f(\varphi)= \dfrac{\alpha'}{16\kappa_D^2} e^{-\kappa_1\varphi}\ ,\quad g(\varphi)={e^{-2\kappa_1\varphi}\over {\kappa_D^2}} \ ,\label{eq:func_form}
\ee
with $\kappa_1=2{\kappa_D}/\sqrt{D-2}$ and $D=10$. The Gauss-Bonnet scalar is defined as
\be
\mathcal{G}= R_{ABCD}R^{ABCD}-4R_{AB}R^{AB}+R^2.
\ee
This is  the combination of the quadratic curvature corrections that emerges in the effective theory for supersymmetric strings after super-completing the multiplet with the Lorentz-Chern-Simons terms, guaranteeing  the absence of ghosts \cite{Zwiebach, Zumino, Cecotti}.  For $V=0$, the action \eqref{eq:EA_camp} describes the low-energy dynamics of the heterotic superstring \cite{Gross}.  Alternatively, by orbifolding $E_8\times E_8$ strings one can obtain a non-supersymmetric theory whose low-energy limit is given by $SO(16)\times SO(16)$ strings, where the vacuum energy generates a potential \cite{Gaume},
\be \label{V}
V(\varphi)=\frac{0.037}{\alpha'^{5}}e^{5\kappa_1\varphi}\ .
\ee
The effective action  \eqref{eq:EA_camp}  is now a consistent truncation of this $SO(16)\times SO(16)$ action. 

In this paper, we shall work with a slight generalisation of  \eqref{eq:EA_camp}, as given by  \eqref{eq:action}, allowing arbitrary spacetime dimension, $D$,  and  a $k$-form that is not necessarily a 3-form. We will also consider more general potentials, focussing on  exponentials with arbitrary coefficients and arbitrary slopes. Actually, these coefficients are not \emph{entirely} arbitrary ---~we will assume they have the same sign as in the known example from the heterotic string \eqref{eq:EA_camp}. This means that all the cases of interest will follow the philosophy of \cite{Montero}, a runaway potential to be lifted by flux contributions. Note that the relative sign of the slopes is all that really matters in the corresponding potentials. This because one can always perform a redefinition of the scalar $\varphi\to-\varphi$. With our conventions, the runaway is assumed to occur in the limit $\varphi\to-\infty$, in contrast with the conventions of \cite{Montero}, but consistent with \cite{Campbell}. 

In the remaining sections, we consider compactified solutions of the action \eqref{eq:action} that are a direct product of $n$-dimensional de Sitter/Minkowski times a $p$-sphere/torus. In terms of the Riemann tensors associated with the external and internal metrics, this translates as
\begin{align}
	\tilde R_{\mu\nu\rho\sigma}&=\kappa\left(\tilde g_{\mu\rho} \tilde g_{\nu\sigma}-\tilde g_{\mu\sigma} \tilde g_{\nu\rho}\right),
	\label{eq:extR}
	\\	
	\hat R_{abcd}&=\lambda\left(\hat g_{ac} \hat g_{bd}-\hat g_{ad} \hat g_{bc}\right),
	\label{eq:intR}
\end{align}
where $\kappa$ is the curvature scale of the physical spacetime, while $\lambda$ is the curvature scale of the internal spacetime ($\lambda>0$ corresponds to a $p$-sphere and $\lambda=0$ to a $p$-torus).  We will also assume that $k$ is either $n$ or $p$, ensuring that the $k$-forms fills either the external or internal space with flux. We remark that in the standard  heterotic case, with $D=10$ and $k=3$ this requires compactification on a 3-sphere or a 3-torus, down to seven external dimensions, requiring three more compact directions to recover a four dimensional universe.  Of course, an explicit embedding of this kind with stable moduli is a non-trivial matter and further consistency considerations, such as tadpole cancellation  \cite{Becker:2005nb}, must be taken into account, but this is beyond the scope of this paper.

Let us conclude by noting that the cases described in this paper do not fall into the usual Maldacena-Nu\~nez no-go theorem \cite{Maldacena} since the Gauss-Bonnet term represents higher curvature corrections to the action that are not considered in the original paper.


\section{Extending the Montero-van Riet-Venken bound in the small Gauss-Bonnet coupling limit}
\label{sec:genpot}

The goal  of this paper is to study the impact of the higher curvature corrections on the MvRV bound for the existence of de Sitter solutions. To develop some intuition, we begin by studying the  Gauss-Bonnet correction using a perturbative expansion in the dimensionless parameter $\epsilon\equiv {\kappa_D^4\langle f(\varphi)/ V(\varphi)\rangle} \geq 0$, where $f(\varphi)$ and $V(\varphi)$ are evaluated at the vev of $\varphi$. As a result, we extend the MvRV bound \eqref{eq:constraint} to account for the higher curvature terms and show that the condition on the existence of de Sitter vacua can be recast in the form
\begin{equation}
    \left|\frac{g'}{g}\right|\geq (p-1)\left|\frac{V'}{V}\right| - \epsilon\cdot h(\varphi,\rho) +\mathcal{O}\left(\epsilon^2\right)\ , \label{eq:Montero_cond_GB}
\end{equation}
where the sign of $h(\varphi,\rho)$ determines whether, at linear order, the Gauss-Bonnet correction favours the existence of solutions or not. In particular, if $h>0$, the bound is more easily satisfied and the higher curvature corrections should make it easier to find de Sitter solutions; if $h<0$, the bound is harder to satisfy and de Sitter solutions are less forthcoming.  

Throughout this section we consider the top form to be purely internal and we keep the sign of the slopes for the potential and the couplings as given in \eqref{eq:func_form}, without specifying the functional form:
\be
\text{sgn}\left({V'\over V}\right) = 1 \ , \qquad \text{sgn}\left({g'\over g}\right) = -1 \ , \qquad \text{sgn}\left({f'\over f}\right) = -1 \ .
\label{eq:genslopes}
\ee
The Lagrangian for the $n$-dimensional theory, after compactifying the $D$-dimensional action \eqref{eq:action} in a $p$-dimensional internal space, is 
\begin{multline}
\mathcal{L} = \sqrt{-\tilde{g}}\left\{ \dfrac{\tilde{R}}{2{\kappa_D^2}} + \dfrac{\hat{R}}{2{\kappa_D^2}} e^{2\left(\alpha-\beta\right)\rho} - e^{2\alpha \rho} V(\varphi) - g(\varphi) Q^2 e^{2(n-1)\alpha\rho}\right.
\\
\left. + f(\varphi)\left[\tilde{\mathcal{G}} e^{-2\alpha\rho}  +2 \tilde{R}\hat{R}e^{-2\beta\rho}+\hat{\mathcal{G}}e^{2\left(\alpha-2\beta\right)\rho}\right] + ...\vphantom{\dfrac{\tilde{R}}{2{\kappa_D^2}}}\right\}\ ,
\label{eq:Vpot}
\end{multline}
where we have omitted the terms involving derivatives of the scalar fields, as we are interested in studying the system at the minimum in the $\left(\varphi,\rho\right)$ directions. Extremising the Lagrangian with respect to the metric and the two fields yields the conditions, 
\begin{align}
\begin{split}
Q^2&={{{e}^{-2n\alpha\rho}}\over g'(\varphi)} \left\{f'(\varphi) \left[ \lambda^2 p(p-1)\left( p-2 \right)  \left( p-3 \right) {{e}^{4
  \left( \alpha-\beta \right) \rho}}\right .\right.
 \\
 &\left.\left.\vphantom{e^{()}} +2\lambda\kappa n(n-1)p(p-1) {{e}^{2  \left( \alpha-\beta
 \right) \rho}}+\kappa^2 n(n-1)(n-2)(n-3)\right]  -{{e}^{4 \alpha \rho}}V' \left( \varphi\right)\right\}\ ,
 \label{eq:Qcond}
 \end{split}
 \\
\begin{split}
0&=\lambda\left\{{\dfrac{1}{\kappa_D^2}}\left(p-1\right)\left(p+n-2\right)e^{2\left(\alpha-\beta\right)\rho}+2\left(p-1\right)f(\varphi)e^{-2\beta\rho}\left[2\kappa\left(n-2\right) n\left(n-1\right)\vphantom{e^{2\left(\alpha-\beta\right)\rho}}\right.\right.
\\
&\quad\left.\left.+\lambda\left(p-2\right)\left(p-3\right)\left(p+2n-4\right)e^{2\left(\alpha-\beta\right)\rho}\right]\vphantom{\dfrac{1}{\kappa_D^2}}\right\} -2{Q^2} g(\varphi)\left(n-1\right)e^{2\left(n-1\right)\alpha\rho}
\\
&\quad -2\kappa^2 n\left(n-1\right)\left(n-2\right)\left(n-3\right)f(\varphi) e^{-2\alpha\rho}- 2V(\varphi) e^{2\alpha\rho} \vphantom{\left\{e^{()}\right\}}\ ,
\label{eq:kcond}
\end{split}
\\
\begin{split}
0&=\kappa \left\{\vphantom{\dfrac{1}{\kappa_D^2}}2(n-1)(n-2) f(\varphi) \left[\kappa (n-3)(n-4)e^{-2\alpha\rho}+2\lambda p(p-1)e^{-2\beta\rho}\right] \right.
\\
&\quad \left.+{\dfrac{1}{\kappa_D^2}}(n-1)(n-2)\vphantom{e^{-2\beta\rho}}\right\}+ \lambda p (p-1)\left[2\lambda(p-2)(p-3) f(\varphi) e^{2\left(\alpha-2\beta\right)\rho} + \dfrac{e^{2\left(\alpha-\beta\right)\rho}}{{\kappa_D^2}}\right]
\\
&\quad- 2 V(\varphi) e^{2\alpha \rho} - 2 g(\varphi) Q^2 e^{2(n-1)\alpha\rho} \vphantom{\dfrac{e^{2\left(\alpha-\beta\right)\rho}}{\kappa_D^2}}\ ,
\label{eq:kappacond}
\end{split}
\end{align}
It is not possible to solve the above system for arbitrary potentials and dimensions. However, we can solve the system order by order in a perturbative Gauss-Bonnet coupling expansion\footnote{Note that Gauss-Bonnet gravity, being quadratic in curvature, typically yields two distinct vacua, one which is perturbative in the coupling, $f$, and one which is non-perturbative. Our analysis here does not capture the latter, which one might have expected to be unstable anyway \cite{GBpaper}.}. To do so, we take the ansatz
\begin{gather}
    Q^2= Q^2_0 + \epsilon\, Q^2_1\ , \quad     \lambda= \lambda_0 + \epsilon \,\lambda_1\ , \quad   \kappa= \kappa_0 + \epsilon\, \kappa_1\ ,\label{eq:ansatz}
\end{gather}
where the zeroth order quantities solve the previous system of equations with $f(\phi)=0$ and are given by
\begin{align}
Q^2_0&={{e}^{-2\left(n-2\right)\alpha\rho}}\left|\frac{{
V'}}{g'  }\right|\ , 
 \label{eq:Q_exp}
\\
{\lambda}_0&={2{\kappa_D^2} V{{e}^{2\beta\rho}}\over (p-1){\left( 
n+p-2\right)}}\left|{g\over g'}\right|\left[\left|{g'\over g}\right|  +\left( n-1
 \right)  \left|{V'\over V}\right|  \right] \ ,
\label{eq:k_exp}
\\
\kappa_0 &= {2{\kappa_D^2}V e^{2\alpha \rho} \over (p+n-2) (n-1)}\left|{g\over g'}\right|\left[\left|{g'\over g}\right|-\left(p-1\right)\left|{V'\over V}\right|\right]\ .
\label{eq:kappa_exp}
\end{align}
Solving for the ansatz \eqref{eq:ansatz} to leading order in the Gauss-Bonnet coupling, we find the external curvature to be
\begin{equation}
\kappa = {2{\kappa_D^2} V e^{2\alpha\rho}\over (p+n-2)(n-1) }\left|{g\over g'}\right|\left[\left|{g'\over g}\right| - (p-1) \left|{V'\over V}\right|+\epsilon h_{n,p}+\mathcal{O}\left(\epsilon^2\right)\right]\ , \label{eq:vmin}
\end{equation}
where $(\varphi,\rho)$ are understood to be evaluated at their vevs. Demanding that the external curvature is positive leads to an extended condition of the form  \eqref{eq:Montero_cond_GB}, with  the leading order correction given by a polynomial in $|g/g'|$
\begin{equation}
\begin{split}
h_{n,p}&={{V^2}\over (p-1)(n-1)(p+n-2)}\left[A_{n,p}\left|{g'\over g}\right|+B_{n,p}\left|{f'\over f}\right|+C_{n,p}\left|{V'\over V}\right|\vphantom{\left|{g'\over g}\right|^2}\right.
\\
&\quad\left.+\left(D_{n,p}\left|{f'\over f}\right|+E_{n,p}\left|{V'\over V}\right|\right)\left|{V'\over V}\right|\left|{g\over g'}\right|+F_{n,p}\left|{f'\over f }\right|\left|{V'\over  V}\right|^2\left|{g\over g'}\right|^2\right]\ .
\label{eq:gcond_app}
\end{split}
\end{equation}
The coefficients of the polynomial depend on the dimensions $\left(n,p\right)$: 
\begin{align}
    A_{n,p}&= 4n(p-7)(1-p) + 4n^2(1-p) + 4p(p-11) + 48 \ ,
    \\
    B_{n,p}&= -4\left[(n - 1)p^2 + (n^2 - 4n + 3)p - n^2 + 3n\right](p - 1)\ ,
    \\
    C_{n,p}&= 8\left[(n -1 )p^3 + (n^2 -11n + 14)p^2 - (2n^2 - 19n + 27)p + n^2 - 7n + 12\right]\ ,
    \\
    D_{n,p}&= 8(p- n)(np - n - p + 3)(p - 1)\ ,
    \\
    E_{n,p}&=4\left[2(3n - 5)p^3 - (2n^2 + 15n - 31)p^2 + (3n^2 + 12n - 31)p + n^2 - 7n + 12\right],
    \\
    F_{n,p}&=4(p - 1)(4n^2p^2 - 7n^2p - 7np^2 + n^2 + 14np + p^2 - 3n - 3p)\ .
\end{align}
Although these  expressions for the coefficients are not very illuminating, we can remark on some interesting behaviour. 

For the particular stringy potentials  given in \eqref{eq:func_form} and with the well motivated  choice,  $n=4$, $p=6$ we find that $h_{n,p}>0$.  However, this does not point to the existence of new de Sitter solutions as the MvRV bound is already strongly violated at leading order.  For the Gauss-Bonnet corrections to yield something interesting in this perturbative set-up, our best hope is ask what happens when the MkRV bound is \emph{only just} violated at leading order.  In particular, we consider the case where
\begin{equation}
\left|\dfrac{g'}{g}\right|=(p-1) \left|\dfrac{V'}{V}\right|(1-\Delta)
\end{equation}
for some $0<\Delta \ll \epsilon$.  In such a scenario, the existence of de Sitter solutions is only marginally ruled out by the original MvRV bound \eqref{eq:constraint}. Could the curvature corrections rule them back in? In this particular case we find that 
\begin{equation}
h_{n,p}=  {4p(p-2)(p-3)\over(p-1)^3}{V^2}\left[\left|\dfrac{g'}{g}\right|-(p-1)\left|\dfrac{f'}{f}\right|\right]  [1+\mathcal{O}(\Delta)] \ , \label{eq:comb}
\end{equation}
Since $0<\Delta \ll \epsilon$, the external curvature is dominated by this form of $h_{n, p}$,  giving
\be
\kappa \approx  \frac{8\kappa_D^2 p(p-2)(p-3) V^3 e^{2\alpha\rho}}{(p+n-2)(n-1)(p-1)^3 }  \epsilon  \left|{g\over g'}\right|
\left[\left|\dfrac{g'}{g}\right|-(p-1)\left|\dfrac{f'}{f}\right|\right] 
\ee
which is clearly positive whenever 
\be
\left|\dfrac{g'}{g}\right| > (p-1)\left|\dfrac{f'}{f}\right|\ . \label{eq:extended_bound}
\ee
This suggest that de Sitter solutions {\it might} be possible for these parametric choices,  although we are not aware of any stringy motivated  supergravity model that satisfies them. 

Another possibility is to consider the case where  $|g'/ g|\ll |V'/V|$.  When this hierarchy is big enough, the last term in \eqref{eq:gcond_app} could be the dominant contribution to the external curvature,  dominating over the zeroth order piece in the perturbative expansion in the Gauss-Bonnet coupling. Whilst this casts some doubt on the validity of the expansion, it would, if true,  suggest that \be
\kappa \approx {2{\kappa_D^2} V^3 e^{2\alpha\rho}\over (p+n-2)^2(n-1)^2(p-1) } \epsilon F_{n,p}\left|{f'\over f }\right|\left|{V'\over  V}\right|^2\left|{g\over g'}\right|^3
\ee
This could point towards new de Sitter solutions  since we also have  that $F_{n,p}>0$ for $n,p\geq2$.   Of course, such an extreme hierarchy is not especially well motivated since generically we expect $|g'/g|$ and $|V'/V| $ to be $\mathcal{O}(1)$ as  both couplings are expected to be fixed by Weyl rescaling to Einstein frame.

\section{Non-perturbative conditions for the existence of de Sitter solutions with a Gauss-Bonnet term}
\label{sec:specifpot}

We shall now establish some exact results, seeking de Sitter and non-trivial Minkowski vacua beyond the perturbative approach of the previous section.  To this end, it is necessary  to fix some of the potentials in order to make concrete progress. We choose the Gauss-Bonnet  potential $f$ and the gauge coupling $g$ to coincide with the heterotic string, as per equation~\eqref{eq:func_form}, so that 
\be \label{fg}
f(\varphi)= \dfrac{\alpha'}{16\kappa_D^2} e^{-\kappa_1\varphi}\ ,\quad g(\varphi)={e^{-2\kappa_1\varphi}\over {\kappa_D^2}} 
\ee
while  the remaining potential is given more generally as  
\begin{equation}
    V(\varphi)=V_0e^{q\kappa_1\varphi},  \label{f}
\end{equation}
with $V_0>0$ and $q>0$ a free dimensionless parameter.  There is sufficient generality left in $V$ to mirror the analysis of \cite{Montero}, and investigate parametric constraints on the slope of $\ln V$. Note that for  the string motivated scenario of \cite{Gaume}, we have $q=5$ and $V_0={0.037\over\alpha'^{5}}$. 

In \cite{Montero}, the slope of $\ln V$ is constrained in the form of the  MvRV bound   \eqref{eq:constraint}. For these exponential potentials, this bound  takes a particularly simple form
\begin{equation}
    q\leq\dfrac{2}{p-1}.
    \label{eq:MvR}
\end{equation}
Our goal here is to ask if the parametric constraint on the slope of $\ln V$ is affected by the Gauss-Bonnet correction. We will also allow for both internal and external fluxes, further generalising \cite {Montero}. If  $H_k$ is allowed to carry external flux, we find that there are only trivial vacua:  a Minkowski external space, vanishing flux and a toroidal internal manifold. If, instead, $H_k$ is allowed to carry internal flux, the set-up is more interesting. The perturbative analysis of the previous section suggested that one could find de Sitter solutions even when the  MvRV  bound is violated, provided an extended condition  \eqref{eq:extended_bound} holds for the Gauss-Bonnet potentials. However, it turns out that this extended bound  is {\it not} satisfied for the string motivated potentials for $f$ and $g$  given by \eqref{fg}.  Indeed, we shall see that in this instance, the MvRV condition \eqref{eq:MvR} is necessary for the existence of de Sitter solutions, but it is not sufficient. This means the higher curvature correction makes de Sitter solutions no more forthcoming, constraining the parametric dependence  of the scalar potential $\ln V$ at least as much as in \cite{Montero}. 

Note that in this section, we shall not assume the particular choice of $\alpha$ and $\beta$ given in \eqref{albe}, but will keep things general. Furthermore,  we work directly with the $D$-dimensional field equations derived from the variation of the action \eqref{eq:action}, which are given by 
\begin{align}
	\begin{split}
		\label{eq:EinsteinD}
		0&=f(\varphi)\left(4R_A^{~C}R_{BC}+\dfrac12\mathcal{G}g_{AB}-2RR_{AB}+4R^{CD}R_{ACBD}-2R_A^{~CDE}R_{BCDE}\right)
		\\
		&\quad+\dfrac12\nabla_A\varphi\nabla_B\varphi-\dfrac14(\nabla\varphi)^2g_{AB}+2f'(\varphi)\left(R\,\nabla_A\nabla_B\varphi-4R_{(B}^{~C}\nabla_{A)}\nabla_C\varphi+2G_{AB}\Box\varphi\right.
		\\
		&\quad+2g_{AB}R^{CD}\nabla_C\nabla_D\varphi-2R_{ACBD}\nabla^C\nabla^D\varphi\left.\vphantom{R_{(B}^{~C}}\right)+2f''(\varphi)\left[R\,\nabla_A\varphi\nabla_B\varphi\right.\vphantom{\dfrac12}
		\\
		&\quad\left.-4R_{C(B}\nabla_{A)}\varphi\nabla^C\varphi+2G_{AB}(\nabla\varphi)^2+2g_{AB}R^{CD}\nabla_B\varphi\nabla_C\varphi-2R_{ACBD}\nabla^C\varphi\nabla^D\varphi\right] \vphantom{\dfrac12}
		\\
		&\quad-\dfrac{1}{2\kappa_D^2}G_{AB}-g(\varphi)\left[\dfrac{1}{2k!}H_{A_1...A_k}H^{A_1...A_k}g_{AB}-\dfrac{1}{(k-1)!}H_{A_1...A_{k-1}A}H^{A_1...A_{k-1}}_{~~~~~~~~~B}\right]
		\\
		&\quad-\dfrac12V(\varphi)g_{AB},
	\end{split}
	\\
	0&=\Box\varphi+f'(\varphi)\mathcal{G}-V'(\varphi)-\dfrac{g'(\varphi)}{k!}H_{A_1...A_k}H^{A_1...A_k},
	\label{eq:scalD}
	\\
	0&=\nabla_B\left[g(\varphi)H^{A_1...A_{k-1}B}\right].
	\label{eq:topform}
\end{align}

\subsection{Internal flux}
\label{ssec:int}
If we assume that $H$ is a top form along the internal manifold, then the solution to equation~\eqref{eq:topform} is given by
\begin{equation}
	H_{a_1...a_p}= Q \hat\epsilon_{a_1...a_p}
	\label{eq:intform}
\end{equation}
where $Q$ is a constant, all other components of $H$ vanish, and $\hat\epsilon_{a_1...a_p}$ is the volume form associated with the internal submanifold ($\hat\epsilon_{1...p}=\sqrt{\hat g}$, etc). We substitute this solution, together with \eqref{eq:extR} and \eqref{eq:intR} into eqs.~\eqref{eq:EinsteinD} and \eqref{eq:scalD}, projecting equation~\eqref{eq:EinsteinD} over $\tilde g_{\mu\nu}$ and $\hat g_{ab}$ respectively. Let us define the following master variables
\begin{equation}
	\label{eq:changevarint}
	\begin{split}
		K&=\kappa e^{2(\beta-\alpha)\rho},\qquad  \hat H=Q^2e^{-2\kappa_1\varphi-2(p-1)\beta\rho},\qquad U=V_0\kappa_D^2e^{q\kappa_1\varphi+2\beta\rho},
		\\
		X_1&=\dfrac{\alpha'}{16} e^{-\kappa_1\varphi-2\beta\rho},\qquad X_2=\dfrac{\alpha'}{16} \kappa e^{-\kappa_1\varphi-2\alpha\rho},\qquad X_3= \dfrac{\alpha'}{16}  \kappa^2 e^{-\kappa_1\varphi+2(\beta-2\alpha)\rho},
	\end{split}
\end{equation}
Given that $\alpha'>0$, for a consistent de Sitter (or Minkowski) vacuum, we necessarily have that all of the master variables, with the possible exception of $U$, are non-negative. This will be crucial to our subsequent proofs. The system of equations can now be cast in the following linear form, which is particularly useful for analysing the system
\begin{align}
	\begin{split}
		\label{eq:Einsteinnint2}
		0&=-2 \hat H - 2 U + (n-1)(n-2) [K + 2 (n-3)(n-4) X_3] 
		\\
		&\quad+ \lambda p(p-1) [1 + 4 (n-1)(n-2) X_2]  +  2\lambda^2 p(p-1)(p-2)(p-3) X_1,
	\end{split}
	\\
	\begin{split}
		\label{eq:Einsteinpint2}
		0&=2 \hat H- 2 U + n(n-1)[K + 2 (n-2)(n-3)X_3] 
		\\
		&\quad+\lambda  (p-1)(p-2)  [1 + 4 n(n-1) X_2] +  2 \lambda^2(p-1)(p-2)(p-3)(p-4) X_1 ,
	\end{split}
	\\
	\begin{split}
		\label{eq:scalint2}
		0&= 2 \hat H - q U - n(n-1)(n-2)(n-3) X_3 - 2\lambda n (n-1) p(p-1) X_2 
		\\
		&\quad- \lambda^2 p(p-1)(p-2)(p-3) X_1 .
	\end{split}
\end{align}
 Given that $n\geq2$ and $q>0$, we can always solve the above equations for $U$, $\hat H$ and $K$ in terms of $X_1$, $X_2$ and $X_3$. In particular, we have 
\begin{equation}
\begin{split}
K=K^\mathrm{(int)}_\mathrm{sol}(X_1,X_2,X_3)&\equiv-\dfrac{1}{(n-1) [2 + (n-1) q]}\left\{\lambda(p-1) [q(p-1)-2]\right.
\\
&\quad+2 \lambda^2(p-1)(p-2)(p-3) [p-4 + (p-2) q] X_1 \vphantom{\dfrac{1}{(n-1) [2 + (n-1) q]}}
\\
&\quad + 4\lambda(p-1) (n-1)\{-p (q-2) + n [p-2 + (p-1) q]\} X_2 \vphantom{\dfrac{1}{(n-1) [2 + (n-1) q]}}
\\
&\quad \left.+2 (n-1)(n-2)(n-3)[4 + n + (n-2) q] X_3\right\}\vphantom{\dfrac{1}{(n-1) [2 + (n-1) q]}}.
\end{split}
\end{equation}
The idea behind the proof is to show that $K$ is maximal at $(X_1,X_2,X_3)=(0,0,0)$ (restricting of course to non-negative $X_i$ values), but that even at its maximum, $K<0$ as soon as we choose $q>2/(p-1)$.  Since $K$ carries the same sign as the external curvature, this would prove the absence of de Sitter solutions whenever $q>2/(p-1)$. Said another way, the MvRV condition \eqref{eq:MvR} is a necessary condition for the existence of de Sitter vacua, even in the presence of  the Gauss-Bonnet correction. 

To this end, we  consider
\begin{align}
\dfrac{\partial K^\mathrm{(int)}_\mathrm{sol}}{\partial X_1}&=-\dfrac{2\lambda^2 (p-1)(p-2)(p-3)[p-4 + (p-2) q]}{(n-1) [2 + (n-1) q]},
\\
\dfrac{\partial K^\mathrm{(int)}_\mathrm{sol}}{\partial X_2}&=-\dfrac{4\lambda (p-1)\{-p (q-2) + n [p-2+ (p-1) q]\}}{2 + (n-1) q},
\\
\dfrac{\partial K^\mathrm{(int)}_\mathrm{sol}}{\partial X_3}&=-\dfrac{2(n-2)(n-3)[n+4+(n-2)q]}{2 + (n-1) q}.
\end{align}
$\partial K^\mathrm{(int)}_\mathrm{sol}/\partial X_3$ is obviously negative. For $p=1,2,3$, $\partial K^\mathrm{(int)}_\mathrm{sol}/\partial X_1=0$ while for $p\geq4$, $\partial K^\mathrm{(int)}_\mathrm{sol}/\partial X_1$ is also obviously negative. Finally, for $p=1$, $\partial K^\mathrm{(int)}_\mathrm{sol}/\partial X_2=0$, while for $p\geq2$ and $n\geq2$,
\begin{equation}
-p (q-2) + n [p-2+ (p-1) q]\geq(p-2)q+4(p-1)>0.
\end{equation}
Hence, $\partial K^\mathrm{(int)}_\mathrm{sol}/\partial X_2\leq0$ in all cases, and $K^\mathrm{(int)}_\mathrm{sol}$ is indeed maximal at $(X_1,X_2,X_3)=(0,0,0)$, where it takes the value
\begin{equation}
K^\mathrm{(int)}_\mathrm{max} \equiv K^\mathrm{(int)}_\mathrm{sol}(0,0,0)=\dfrac{\lambda(p-1) [2-(p-1)q]}{(n-1) [2 + (n-1) q]}.
\label{eq:Kmax}
\end{equation}
The above equation shows that $K$ can vanish for $\lambda=0$ ($p$-torus) or $p=1$ (1-sphere), i.e. when the curvature of the internal manifold vanishes. These solutions correspond to a Minkowski spacetime, and exist only for a vanishing flux $Q$ and a vanishing potential $V(\varphi)$, which makes them uninteresting. The existence of a de Sitter solution, on the other hand, requires that $K>0$. This is not possible if $q>2/(p-1)$, as shown by equation~\eqref{eq:Kmax}, which completes our proof. The condition \eqref{eq:MvR} is necessary for de Sitter solutions to exist.

In terms of the variables $K$, $\hat H$, $U$, $X_i$, \eqref{eq:MvR} is also sufficient; indeed, it is possible to show that, when $q\leq2/(p-1)$, $K$, $\hat H$ and $U$ are all positive in the neighbourhood of the point $(X_1,X_2,X_3)=(0,0,0)$. However, this is an artifact of the choice of variables. Sending all $X_i$ to 0 actually corresponds to sending the Gauss-Bonnet coupling to 0. Hence, it is not surprising to find that condition \eqref{eq:MvR} is necessary and sufficient in this case, as we are back to the action studied in \cite{Montero}. If instead, we think of the Gauss-Bonnet coupling as fixed, we can show that the condition \eqref{eq:MvR} is not sufficient, by an analysis that is very similar to the one we carried out in section \ref{sec:genpot}. Indeed, let us saturate the bound \eqref{eq:MvR} by choosing $q=2/(p-1)$ ---~which corresponds to being on the edge of the existence region when the Gauss-Bonnet coupling is absent. Then, $K_\text{sol}^{(\text{int})}$ becomes
\begin{equation}
\begin{split}
K_\text{sol}^{(\text{int})} &=-\dfrac{1}{(n-1)(n+p-2)}\left\{(n-1) (n-2)(n-3) [n-8 + (4 + n) p] X_2\vphantom{(p-3)^2}\right.
\\
&\quad\left.+ 2p(p-1)(n-1) [(2 + n) p-n-4] \lambda X_3 + p(p-1)(p-2)(p-3)^2 \lambda^2 X_1\right\}.
        \label{eq:Ksolintsat}
        \end{split}
\end{equation}
It is easy to check that for $p\geq2$ (we already saw that the case $p=1$ was uninteresting), the coefficients of $X_1$, $X_2$ and $X_3$ in the above equation are all negative. The coefficient of $X_3$ is actually strictly negative. Hence, there can be no de Sitter (or Minkowski) solution for this value of $q$ ---~and neighbouring ones, by continuity. Therefore, \eqref{eq:MvR} is not sufficient. This reinforces the conclusion of section \ref{sec:genpot}, as it is a non-perturbative statement. The presence of the Gauss-Bonnet term never enlarges the space of solutions, but instead reduces it, at least when the potentials, $f$ and $g$, take on the form best motivated by string theory.

\subsection{External flux}

So far, we have assumed $H$ to be an  internal flux, and,  as we have demonstrated, this leads to the absence of de Sitter solutions to the effective action~\eqref{eq:action} as long as $q>2/(p-1)$. Now let us consider the opposite scenario: that $H$ is a top form along the external directions. In this case, the solution to equation~\eqref{eq:topform} is given by
\begin{equation}
	H_{\mu_1...\mu_n}= \dfrac{Q}{g(\varphi)} e^{(n\alpha-p\beta)\rho} \tilde\epsilon_{\mu_1...\mu_n}
	\label{eq:extform}
\end{equation}
where $Q$ is a constant, all other components of $H$ vanish, and $\tilde\epsilon_{\mu_1...\mu_n}$ is the volume form associated with the internal submanifold ($\tilde\epsilon_{1...n}=\sqrt{-\tilde g}$, etc). We will prove that no de Sitter solutions at all exist in this case, for any value of $q$. The proof is very similar to the one presented in section~\ref{ssec:int}. Employing the same variables $U$, $K$, $X_i$ as defined in equation~\eqref{eq:changevarint}, and trading $\hat H$ for
\begin{equation}
\tilde H =Q^2e^{2\kappa_1\varphi-2(p-1)\beta\rho},
\end{equation}
the field equations are identical to eqs.~\eqref{eq:Einsteinnint2}--\eqref{eq:scalint2} upon the substitution $\hat H\to\tilde H$ in eqs.~\eqref{eq:Einsteinnint2}-\eqref{eq:Einsteinpint2} and $\hat H\to-\tilde H$ in equation~\eqref{eq:scalint2}. We need to distinguish a few more sub-cases with respect to the internal case, depending on how $q$ compares to $2/(n-1)$.

\subsubsection{$q>2/(n-1)$}

In this case, the proof is almost exactly identical to the internal case. We solve the field equations for $K$, $U$ and $\tilde H$ in terms of $X_1$, $X_2$ and $X_3$, to obtain
\begin{equation}
\begin{split}
K=K^\mathrm{(ext)}_\mathrm{sol}(X_1,X_2,X_3)&\equiv-\dfrac{1}{(n-1) [(n-1) q-2]}\left\{\lambda(p-1) [q(p-1)+2]\right.
\\
&\quad+2 \lambda^2(p-1)(p-2)(p-3) [p+4 + (p-2) q] X_1\vphantom{\dfrac{1}{(n-1) [2 + (n-1) q]}}
\\
&\quad+4 \lambda(n-1)(p-1)\{-p (q+2) + n [p+2 + (p-1) q]\} X_2 \vphantom{\dfrac{1}{(n-1) [2 + (n-1) q]}}
\\
&\quad \left.  +2 (n-1)(n-2)(n-3)[n-4 + (n-2) q] X_3\right\}\vphantom{\dfrac{1}{(n-1) [(n-1) q-2]}}.
\end{split}
\end{equation}
It immediately follows that
\begin{align}
\dfrac{\partial K^\mathrm{(ext)}_\mathrm{sol}}{\partial X_1}&=-\dfrac{2\lambda^2 (p-1)(p-2)(p-3)[p+4 + (p-2) q]}{(n-1) [(n-1) q-2]},
\\
\dfrac{\partial K^\mathrm{(ext)}_\mathrm{sol}}{\partial X_2}&=-\dfrac{4\lambda (p-1)\{-p (q+2) + n [p+2+ (p-1) q]\}}{(n-1) q-2},
\\
\dfrac{\partial K^\mathrm{(ext)}_\mathrm{sol}}{\partial X_3}&=-\dfrac{2(n-2)(n-3)[n-4+(n-2)q]}{(n-1) q-2}.
\end{align}
Since $q>2/(n-1)$, $\partial K^\mathrm{(ext)}_\mathrm{sol}/\partial X_1$ is obviously negative. For $n=2,3$, $\partial K^\mathrm{(ext)}_\mathrm{sol}/\partial X_3=0$ while for $n\geq4$, $\partial K^\mathrm{(ext)}_\mathrm{sol}/\partial X_3$ is also obviously negative. Finally, for $p=1$, $\partial K^\mathrm{(ext)}_\mathrm{sol}/\partial X_2=0$, while for $p\geq2$ and $n\geq2$,
\begin{equation}
-p (q+2) + n [p+2+ (p-1) q]\geq(p-2)q+4>0.
\end{equation}
Hence, $\partial K^\mathrm{(ext)}_\mathrm{sol}/\partial X_2\leq0$ in all cases, and $K^\mathrm{(ext)}_\mathrm{sol}$ is maximal at $(X_1,X_2,X_3)=(0,0,0)$, where it takes the value
\begin{equation}
K^\mathrm{(ext)}_\mathrm{max} \equiv K^\mathrm{(ext)}_\mathrm{sol}(0,0,0)=-\dfrac{\lambda(p-1) [2+(p-1)q]}{(n-1) [(n-1) q-2]}.
\end{equation}
This is always negative in the sub-case that we are considering, hence no de Sitter solutions exist. Again, trivial Minkowski solutions can exist for $p=1$ or $\lambda=0$.

\subsubsection{$q<2/(n-1)$}

Once again, the proof is very analogous to the previous ones, but now we consider the solution for $\tilde H$ rather than $K$. This is given by
\begin{equation}
\begin{split}
\tilde H=\tilde H_\mathrm{sol}(X_1,X_2,X_3)&\equiv\dfrac{1}{2[(n-1) q-2]}\left\{\lambda(p-1) q(n+p-2)\right.
\\
&\quad+2 \lambda^2(p-1)(p-2)(p-3) [p + (2n+p-4) q] X_1\vphantom{\dfrac{1}{(n-1) [2 + (n-1) q]}}
\\
&\quad+4 \lambda n(n-1)(p-1)[p + (n-2) q] X_2 \vphantom{\dfrac{1}{(n-1) [2 + (n-1) q]}}
\\
&\quad \left.  -2 n(n-1)(n-2)(n-3)(q-1) X_3\right\}\vphantom{\dfrac{1}{(n-1) [(n-1) q-2]}}.
\end{split}
\end{equation}
and therefore,
\begin{align}
\dfrac{\partial \tilde H_\mathrm{sol}}{\partial X_1}&=\dfrac{2\lambda^2 (p-1)(p-2)(p-3)[p+(p+2n-4)q]}{(n-1) [(n-1) q-2]},
\\
\dfrac{\partial \tilde H_\mathrm{sol}}{\partial X_2}&=\dfrac{4\lambda n(n-1)(p-1)[p+(n-2)q]}{(n-1) q-2},
\\
\dfrac{\partial \tilde H_\mathrm{sol}}{\partial X_3}&=\dfrac{n(n-1)(n-2)(n-3)(1-q)}{(n-1) q-2}.
\end{align}
$\partial \tilde H_\mathrm{sol}/\partial X_1$ and $\partial \tilde H_\mathrm{sol}/\partial X_2$ are obviously negative in the subcase we consider. For $n\leq3$, $\partial \tilde H_\mathrm{sol}/\partial X_3=0$. For $n\geq4$, since $q<2/(n-1)$, $1-q>0$ and therefore $\partial \tilde H_\mathrm{sol}/\partial X_3$ is negative. Hence,  $\tilde H_\mathrm{sol}$ is maximal at $(X_1,X_2,X_3)=(0,0,0)$, where it takes the value
\begin{equation}
\tilde H_\mathrm{max} \equiv \tilde H_\mathrm{sol}(0,0,0)=\dfrac{\lambda q(p-1) (n+p-2)}{2 [(n-1) q-2]}<0.
\end{equation}
Since $\tilde H$ should be non-negative for real values of the flux, we conclude that  the underlying assumptions are inconsistent and therefore no de Sitter or non-trivial Minkowski solutions can exist.

\subsubsection{$q=2/(n-1)$}

In the case where $q$ takes exactly the value $2/(n-1)$, we can no longer solve the system of equations for $K$, $U$ and $\tilde H$ simultaneously. However, we can solve it for $X_2$, $U$ and $\tilde H$ instead, provided that $\lambda\neq0$ and $p\geq2$ (if this is not the case, once again only trivial Minkowski solutions can exist). We obtain
\begin{equation}
\begin{split}
X_2=X_{2,\,\mathrm{sol}}(X_1,X_3)&\equiv-\dfrac{1}{2\lambda n(n-1) (p-1) [(n-1)p+2(n-2)]}
\\
&\quad\left\{ \lambda^2 (p-1)(p-2)(p-3) [p(n+1)+4(n-2)] X_1 \right.\vphantom{\dfrac{1}{(n-1) [2 + (n-1) q]}}
\\
&\quad \left. +n(n-1)(n-2)(n-3)^2 X_3 +\lambda(p-1) (n + p-2)\right\}\vphantom{\dfrac{1}{(n-1) [(n-1) q-2]}}.
\end{split}
\end{equation}
Hence,
\begin{align}
\dfrac{\partial \tilde X_{2,\,\mathrm{sol}}}{\partial X_1}&=-\dfrac{\lambda(p-2)(p-3)[p(n+1)+4(n-2)]}{2 n(n-1)[(n-1)p+2(n-2)]},
\\
\dfrac{\partial \tilde X_{2,\,\mathrm{sol}}}{\partial X_3}&=-\dfrac{(n-2)(n-3)^2}{2\lambda (p-1) [(n-1)p+2(n-2)]},
\end{align}
These quantities are obviously negative. Therefore,  $X_{2,\,\mathrm{sol}}$ is maximal at $(X_1,X_3)=(0,0)$, where it takes the value
\begin{equation}
X_{2,\,\mathrm{max}} \equiv \tilde X_{2,\,\mathrm{sol}}(0,0)=-\dfrac{n+p-2}{2 n(n-1) [(n-1)p+2(n-2)]}<0.
\end{equation}
Again, for consistent de Sitter or non-trivial Minkowski solultions, $X_2$ should be non-negative, in contradiction with the above inequality. This concludes the proof for the external case.

\section{Discussion}
\label{sec:discussion}
We have considered solutions to a general class of gravitational actions, motivated by quadratic curvature corrections to the effective action for the heterotic string.  The actions include a scalar field (identified with the dilaton), a $k$-form field strength, and gravity. We considered warped compactifications on spheres and tori, with $k$-form flux and investigated the existence of de Sitter solutions along the external directions.  A similar analysis was carried out in \cite{Montero} without the quadratic curvature corrections, which in our case corresponded to the Gauss-Bonnet operator coupled to the dilaton.  In \cite{Montero}, the MvRV bound \eqref{eq:constraint} was derived for the dilaton potentials, specifiying the condition for de Sitter solutions to exist. In the presence of the quadratic  curvature corrections, we saw that this bound could be violated  perturbatively and de Sitter solutions could still exist, as long as a second bound \eqref{eq:extended_bound} on the Gauss-Bonnet coupling  also held.  However, for potentials that are best motivated by string theory, the second bound does not hold, suggesting no violation of the MvRV bound is possible in this instance.  Indeed, for these well motivated set-ups, we were also able to show that the MvRV bound was a necessary, but not sufficient, condition for the existence of de Sitter solutions in the presence of quadratic curvature corrections. This leads to our main conclusion:  higher curvature corrections that are well motivated by string theory generically made it parametrically harder, not easier, to find de Sitter solutions. 

It would be very interesting to ask if this were a generic feature of higher order  corrections in all versions of string theory. Clearly this is difficult to establish and more evidence needs to be found.  For  type II strings, curvature corrections kick in at fourth order \cite{GW}, while for heterotic strings we may also consider higher order mixings between the gauge field and curvature \cite{Gross}.  

We can make a crude and simplistic argument to support a claim that curvature  corrections are unlikely to improve the search for de Sitter vacua, at least at finite order in the curvature expansion.  In $D$ dimensional General Relativity without sources, the absence of de Sitter vacua can roughly be understood in terms of an energetic balance between the curvature of the external space and that of the compact space. If the compact space carries positive or vanishing curvature (as in a sphere or a torus), then the external space looks to counter that, delivering negative or vanishing curvature (as in AdS or Minkowski).  For higher curvature operators to 
facilitate the existence of new de Sitter vacua, the details of this balance have to change. In particular, on one side of the energetic balance, a positive curvature space (be it a sphere along the compact directions, or de Sitter along the external directions)  must contribute negative energy through the higher order operators. This might indicate the presence of ghost-like instabilities and the associated negative energy excitations. Such instabilities should certainly be absent in effective actions derived from string theory.

 Of course, the situation becomes more complicated in the presence of additional fields  and sources, including branes, fluxes, Casimir energies, and with more exotic compact spaces.  The challenge here is to retain calculational control over the effective description, be it parametrically or even  just numerically.  Indeed, the level of control one is willing to sacrifice seems to be at the core of the ongoing debate over de Sitter space and the \emph{Swampland}. 
 
\acknowledgments

 We would like to thank  Michele Cicoli, Francisco Pedro and Thomas Van Riet for useful discussions.  The work of AP was funded by STFC Consolidated Grant Number ST/T000732/1. The work of FC was funded by a University of Nottingham  studentship. WTE is supported by the Czech Science Foundation GA\v{C}R, project 20-16531Y. This project has received funding from the European Union's Horizon 2020 research and innovation programme under the Marie Sklodowska-Curie grant agreement No 101007855.
AL thanks FCT for financial support through Project~No.~UIDB/00099/2020.
AL acknowledges financial support provided by FCT/Portugal through grants PTDC/MAT-APL/30043/2017 and PTDC/FIS-AST/7002/2020.
AL would like to acknowledge networking support by the GWverse COST Action
CA16104, ``Black holes, gravitational waves and fundamental physics.''
%



\end{document}